\documentclass[10pt]{iopart}

\newcommand{\mic}{~{\rm \mu m}}

\usepackage{iopams}  
\expandafter\let\csname equation*\endcsname\relax
\expandafter\let\csname endequation*\endcsname\relax
\usepackage{amsmath,amstext,amssymb}
\usepackage{graphicx}
\usepackage{multirow}
\begin{document}

\title[Ultrafast nonlinear dynamics of thin gold films ]{Ultrafast nonlinear dynamics of thin gold films due to an intrinsic delayed nonlinearity}

\author{Morten Bache, Andrei V. Lavrinenko}

\address{Technical University of Denmark, Oersteds Plads, Bld.343, Kongens Lyngby, DK-2800, Denmark}
\ead{alav@fotonik.dtu.dk}
\begin{abstract}
Using long-range surface plasmon polaritons light can propagate in metal nano-scale waveguides for ultracompact opto-electronic devices. Gold is an important material for plasmonic waveguides, but although its linear optical properties are fairly well understood, the nonlinear response is still under investigation. We consider propagation of pulses in ultrathin gold strip waveguides, modeled by the nonlinear Schr\"odinger equation. The nonlinear response of gold is accounted for by the two-temperature model, revealing it as a delayed nonlinearity intrinsic in gold. 
The consequence is that the measured nonlinearities are strongly dependent on pulse duration. This issue has so far only been addressed phenomenologically, but we provide an accurate estimate of the quantitative connection as well as a phenomenological theory to understand the enhanced nonlinear response as the gold thickness is reduced. 
In comparison with the previous works, the analytical model for the power-loss equation has been improved, and can be applied now to cases with a high laser peak power. We show new fits to experimental data from literature and provide updated values for the real and imaginary part of the nonlinear susceptibility of gold for various pulse durations and gold layer thicknesses. Our simulations show that the nonlinear loss is inhibiting efficient nonlinear interaction with low-power laser pulses. We therefore propose to design waveguides suitable for the mid-IR, where the ponderomotive instantaneous nonlinearity can dominate over the delayed hot-electron nonlinearity and provide a suitable plasmonics platform for efficient ultrafast nonlinear optics.
\end{abstract}


\ioptwocol

\section{Introduction}
Plasmonics was aimed for synergy between electronics and photonics, and actually in its current state it offers multiple solutions for optical nanoscale circuitry. In this way it represents a competitive on-chip platform for many optical and photonic applications \cite{Brongersma2010,Bozhevolnyi-2009-book}. Typically plasmonics is subdivided into two main directions. One is dealing with waveguiding on metal-dielectric interfaces of different shapes, and  thus are classified relevant surface plasmon-polaritons (SPPs): gap plasmons, groove plasmons, wedge plasmons, etc. Among an extended nomenclature of plasmonic waveguides, there are cases aimed for integrated optics \cite{Bozhevolnyi-2009-book,Krasavin2015}, nanofocusing \cite{Gramotnev2010,Zenin2015}, sensing \cite{Krupin2013}, light amplification \cite{Berini2012} and even quantum optics. Another direction deals with localized resonances in metal nanoparticles \cite{Maier-book}. Strong field enhancement associated with localized surface plasmons is a natural outcome of any electromagnetic resonance, and thus they facilitate all processes, where field amplitudes or intensities have the dominant role: chemi- and bio-sensing, photovoltaics, absorption, heating, etc. 

However, the role of plasmons either propagating or localized has not been restricted to linear processes only. Every field enhancement happened with involvement of plasmonic materials can be naturally connected to optical nonlinearities. There are a lot of examples reported for plasmonics-enabled nonlinearities \cite{Kauranen2012}. Here we should note that there are two types of effective nonlinearities associated with plasmonic elements, like waveguides, particles, etc. The first one is nonlinearities brought by traditional nonlinear optical materials, but enhancement being supported by plasmonic behavior of auxiliary elements. Waveguides, especially hybrid plasmonic waveguides, arrays of plasmonic nanoparticles and other configurations as parts of future nanophotonic communications contours have been probed for such performance \cite{Kauranen2012,Diaz2016a,Diaz2016}. The second type is explicitly connected with the intrinsic nonlinearities of metal or any other plasmonic-like materials, e.g. transparent conductive oxides. Bulk metals, thin metal layers, and plasmonic metamaterials have been investigated in the nonlinear regime \cite{Kauranen2012, Ginzburg2013,Neira2015,baron,Lysenko2016,Lysenko2016b}.  

In the frames of the second direction aiming for communication applications with nonlinearities originated from metals, gold so far exhibits superior behavior, having simultaneously decent plasmonic performance and fabrication-wise flexibility. On top of that, it is one of the most extensively studied metals in photonics and yet the role of the nonlinear response is still not fully understood. Our motivation is to get a better understanding of how gold as a prototypical metal behaves when illuminated with an ultrashort laser pulse (fs-ps regime) with enough power to observe nonlinear effects, and use this understanding to generalize to other metals. 
We will focus mostly on the third-order nonlinearity of gold as considered to be most encountered case, see for example recent analysis of broadly-dispersed $\chi^{(3)}$ data for gold by Boyd et al. \cite{Boyd2014}. 

Contributions to the third-order nonlinear susceptibility are thoroughly discussed in \cite{Hache:1988}. Accordingly to Hache et al. the dominating role is with the interband transitions and hot-electron contribution. The input from the intraband transitions is few orders of magnitude smaller than the other two and thus can be neglected. Among these two parts, thermomodulated changes to the interband  transitions prevail in noble metals. To be noted, in later works, for example \cite{Marini2013} the delayed response of hot electrons is considered as addition to the interband dielectric function, thus having effectively at the end only the interband and intraband contributions to $\chi^{(3)}$ of gold.

As calculated by Marini et al. \cite{Marini2013} the most pronounced changes for gold nonlinear susceptibility happen in the wavelength range of 400-600 nm, consequently, most of the data analyzed in \cite{Boyd2014} were collected here. Typically there were revealed after the z-scan experiments reporting very high values of $\chi^{(3)}\sim 10^{-16}$-$10^{-15}\, \rm m^2/V^2$ \cite{Smith1999,Lee2006,Xenogiannopoulou2007,Rotenberg-PhysRevB.75.155426}. In contrary, communication-driven nanophotonic and plasmonic devices are extensively exploited in the infrared (IR) range  \cite{Bozhevolnyi-2009-book,Krasavin2015}, where the propagation losses of long-range SPPs (LRSPPs) in gold-based waveguides at the telecommunication wavelengths are well-ameliorated down to $\sim$0.6 dB/mm  \cite{kinsey:2015}. However, the third-order nonlinear susceptibility
of gold in the IR range is much smaller than in the visible \cite{Neira2015}. On the other hand, $\chi^{(3)}$ is still large enough to exhibit detectable effects of nonlinear propagation in plasmonic waveguides, which are supported by field localization near the metal interfaces and  long propagation distances of LRSPPs \cite{Lysenko2016,Lysenko2016b}. 

As the two-temperature model (TTM) is accounted as the basic model for gold nonlinearities \cite{Conforti2013,Marini2013}, there is inevitably a strong pulse duration dependence of the nonlinearities when pulses are compared with the characteristic times of the TTM for gold: thermalization time for hot electrons $\tau_{\rm th} = 300 $ fs and relaxation time of the thermalized electrons to the lattice temperature $\tau_{\rm r} = 1$ ps \cite{Marini2013}. Indeed, such dependence is observed in experiments with the 200 fs and 3 ps laser pulses employed in \cite{Lysenko2016,Lysenko2016b}. A more quantitative interpretation of such dependence was suggested in \cite{Lysenko2016b}. It relied on the TTM as a basis for implementing the nonlinear response of gold in a generalized nonlinear Schr\"odinger equation, as outlined also by Marini et al. \cite{Marini2013}. An important point is that the delayed nonlinearity gives rise to a quantifiable correction factor of the \textit{observable} nonlinear response for ultrashort pulses, which allows to compare the theoretical values in the cw limit with the values measured with a pulsed laser. 

Experimental results reported in \cite{Lysenko2016,Lysenko2016b} also reveal a  thickness dependence of the hot-electron third-order nonlinear susceptibility. We argued that it could be understood as a direct consequence of the TTM, where the thermo-modulational interband nonlinear susceptibility in the cw pulse duration limit for bulk gold is given by the expression
\begin{equation}
\label{eq:chi3_T}
\chi^{(3)}_{T,\rm cw}(\omega_{0})=\frac{1}{2}\varepsilon_{0}\omega_{0}{\rm Im}(\varepsilon_{m}(\omega_{0}))\gamma_T(\omega_{0}),
\end{equation}
where ${\rm Im}(\varepsilon_{m})$ is responsible for the absorption power and $\gamma_T(\omega)$ is a function defined in \cite{Marini2013} through some empirical constants and the temperature derivative of the dielectric function of gold. 
The increased nonlinear response when the gold layer thickness becomes nanometer-scale can then be modelled phenomenologically by suggesting that the imaginary part of the dielectric function of gold become inversely proportional to the thickness of the gold layer. A $1/R$ size-dependent behavior of the collision frequency in the Drude model of the permittivity in metals is well-known in application to nanoparticles with radius $R$ \cite{kreibig1985,kreibig1995}. In turn, it has not been typically applied in studies of LRSPP propagation in ultrathin metal strips. However, such dependence can be attributed to the grain sizes, which for the thin films are constrained by the smallest size parameter, i.e. the thickness. Alternatively, one can think on the size dependence as it reflects the ratio of surface scattering probability (being proportional to the surface area) and the number of electrons, which is volume proportional. Transferring this concept to a film or a strip immediately provides the inverse thickness dependence of the collision frequency. One way or another, the correction to the Drude permittivity should be taken into account. Our hypothesis was confirmed by characterization of nonlinear susceptibility of gold strip waveguides \cite{Lysenko2016,Lysenko2016b}, where the inverse thickness dependence was clearly observed in the analysis of data for three thicknesses of metal strips: 22 nm, 27 nm and 35 nm.


Here we present an in-depth theoretical analysis of the nonlinear Schr\"odinger equation, and show an improved data treatment of the experimental results from \cite{Lysenko2016,Lysenko2016b}. We also show full-scale modelling of short pulse propagation in gold plasmonic waveguides according to the nonlinear Schr\"odinger equation. This was done to get a better understanding on how to interpret the experimental data when trying to evaluate the material parameters for the nonlinear response of gold. 

Ultimately, we also aim to understand whether thin gold waveguides can be used for efficient ultrafast nonlinear optics applications. The nonlinear coefficients of noble metals are 5-6 orders of magnitude above those of standard dielectric materials. So the idea is to excite LRSPPs  with low enough linear propagation losses being able to propagate through a millimeter-long waveguide, long enough to allow nonlinear effects to happen. Our simulations illuminate the potential of such systems, and show that the strong nonlinear loss associated with the delayed hot-electron nonlinearity is a critical obstacle for efficient nonlinear optics with ultrashort laser pulses having low peak powers. 

\section{Ultrafast nonlinear dynamics of gold}

Recent works \cite{Marini2013,Lysenko2016b} have introduced the notion that for a LRSPP propagating in a metal-dielectric waveguide, the ultrafast nonlinear behavior can be modeled by a nonlinear Schr{\"o}dinger equation (NLSE). In the NLSE the cubic nonlinearity has contributions from the \textit{hot electrons} of a metal strip that give a delayed nonlinear self-action on the pulse. In turn, the mode will also substantially overlap into the cladding dielectric materials, here tantalum pentoxide (TaO$_5$) and silica (SiO$_2$). As we outline in \ref{sec:vector}, specifically Eq. (\ref{eq:NLSE_Au}), the following NLSE is appropriate in the co-moving frame
\begin{align}\label{eq:NLSE_Au_TTM}
\left(i\partial_{\zeta}+\hat D_\tau +i\tfrac{1}{2}\alpha\right) A+\hat S (\Gamma_{\rm TaO_5}+\Gamma_{\rm SiO_2}) A|A|^2
\\\nonumber
+\hat S \Gamma_{\rm Au}A\int_{-\infty}^\infty d\tau' h_T(\tau-\tau') |A(\zeta,\tau')|^2=0
\end{align}
with $h_T$ representing the delayed response of the build-up of thermalized hot electrons. The other parameters are the usual ones:  envelope $A$ is normalized so $|A|^2$ gives the instantaneous power in watts, $\hat D_\tau$ is the dispersion operator, $\alpha$ is the linear propagation loss, and $\hat S$ is the self-steepening operator. The nonlinear waveguide coefficients have the usual form 
\begin{align}
\label{eq:Gamma}
\Gamma_{j}&=\frac{3\omega_0 \theta_j \chi^{(3)}_{j}}{4\varepsilon_0 c^2 \tilde n^2 A_{\rm eff}}
, \quad {j=\rm Au,\, TaO_5, \,SiO_2}
\end{align}
where $\tilde n$ is a \textit{generalized} effective index, cf. Eq. (\ref{eq:ntilde}), $A_{\rm eff}$ is the effective waveguide mode area, cf. Eq. (\ref{eq:Aeff}), and $\theta_j$ are the dimensionless nonlinear LRSPP mode field overlap of the various constituent materials of the waveguide, cf. Eq. (\ref{eq:theta-simple}). 

In the quasi-cw limit, the convolution integral of the delayed gold nonlinearity is simplified so the NLSE has the usual form $\left(i\partial_{\zeta}+\hat D_\tau +i\tfrac{1}{2}\alpha\right) A+ \left( 
\gamma_{\rm NL}^{
\rm cw}+i\tfrac{1}{2}\beta_{\rm NL}^{\rm cw} \right) A|A|^2=0$, where 
\begin{multline}
\label{eq:gamma-beta-cw}
\gamma_{\rm NL}^{\rm cw}+i\tfrac{1}{2}\beta_{\rm NL}^{\rm cw}=\frac{3\omega_0}{4\varepsilon_0 c^2 \tilde  n^2 A_{\rm eff}} \\\times 
[\theta_{\rm Au}\chi^{(3)}_{\rm Au}+\theta_{\rm TaO_5}\chi^{(3)}_{\rm TaO_5}+\theta_{\rm SiO_2}\chi^{(3)}_{\rm SiO_2}]
\end{multline}
We here constrict the nonlinearity of gold to the thermo-modulational nonlinearity, Eq. (\ref{eq:chi3_T}). For ultra-thin gold layers we additionally impose a correction by a phenomenological description of the thickness-dependent enhancement that was observed in \cite{Lysenko2016,Lysenko2016b}
\begin{align}
\chi_{\rm Au} ^{(3)}=\chi_{T,{\rm cw}} ^{(3)}(\omega_0)\left(1+\frac{v_F/ \gamma_{f,\infty} }{ t_{\rm Au} }\right)
\label{eq:chi3_cw}
\end{align}
It is represented as the ratio of the characteristic scattering length scale of the the electrons (the ratio between the Fermi velocity of gold $v_F$ and the collision rate of bulk gold $\gamma_{f,\infty}$) and the nanometer thickness of the metal waveguide $t_{\rm Au}$, see the Appendix for more details. To a good approximation the hot-electron dynamics in Eq. (\ref{eq:NLSE_Au_TTM}) is  dominant over other nonlinear effects in the visible and near-infrared. The ponderomotive effect \cite{Ginzburg2010}, cf. Eq. (\ref{eq:chi3_PM}), would give a real and instantaneous contribution, but since it scales as $1/\omega^4$ it is insignificant at high frequencies, and we checked that we can safely neglect it for the cases presented here. In turn it must be included further into the infrared. The cladding material(s) will also contribute with instantaneous and delayed (Raman) nonlinearities, and for simplicity we neglect here Raman effects in the cladding materials. The \textit{effective} contribution $\theta_j \chi_j^{(3)}$ from the cladding materials is often can be neglected. However, below we also show that for the ultrashort pulses the claddings nonlinearity starts to dominate over the hot-electron contribution and must be taken into account.

\begin{figure}[bt]
\centering
\includegraphics[width=\linewidth]{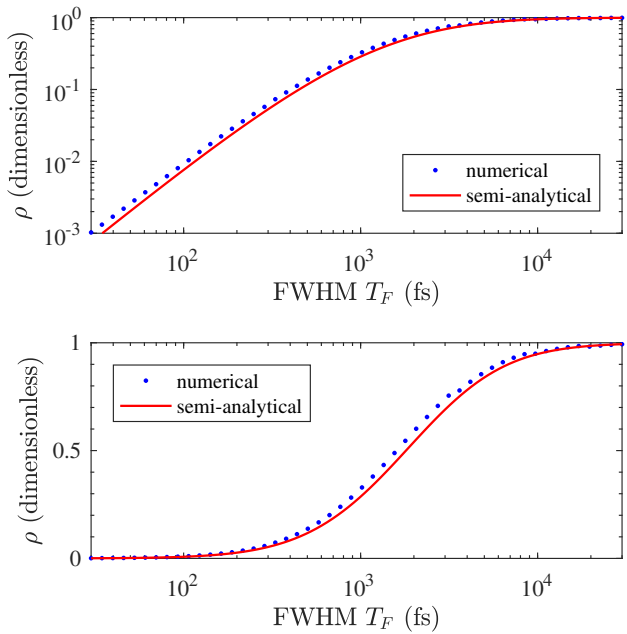}
\caption{Comparing the correction factor as calculated by the semi-analytical approach and that of a full-scale numerical simulation of the NLSE. 
The numerical sweep was performed using $\lambda=1.064 \mic$, where for each pulse duration a series of pulse energies between 0.1-2 nJ were propagated in a 0.2 mm waveguide and the normalized integrated power vs input integrated power was fitted to a polylogarithmic function, cf. Eq. (\ref{eq:TPA-avg-ratio}). The  waveguide nonlinearities were those determined  experimentally at 1064 nm, cf. Table \ref{tab:data}. 
}
\label{fig:rho_numerical}
\end{figure}

In what follows we will mostly study the imaginary part of the nonlinearity, leading to nonlinear losses, and in this particular case the only contribution to the cubic nonlinearity is the delayed hot-electron dynamics, without any approximations, as the dielectric cladding materials and the ponderomotive force only give contributions to the real part. We have recently demonstrated that because the nonlinearity of gold is entirely of a delayed nature, the \textit{observed} nonlinear strength (i.e. the parameter that is extracted from a measurement) is strongly dependent on pump pulse duration $T_0$, so we have to correct for this if we want to get the nonlinear strength in the cw limit, as given by e.g., Eq. (\ref{eq:chi3_T}). The relative strength as the pulse duration is varied can  be accurately expressed by the so-called correction factor $\rho(T_0)$ that is used to gauge the strength of the overlap integral in Eq. (\ref{eq:NLSE_Au_TTM}). The correction factor can be evaluated semi-analytically, c.f. (\ref{eq:rho_semianalytical}) \cite{Lysenko2016b} and shown in Fig.  \ref{fig:rho_numerical}. With this approximation, the evolution of the power can be described by the ordinary differential equation [see Eq. (\ref{eq:NLSE-int-long-pulse-limit-rho})]
\begin{align}\label{eq:NLSE-int-long-pulse-limit-rho-main}
\frac{\partial P}{\partial \zeta}+\alpha P+\beta_{\rm NL}^{\rm obs} P^2=0,
\end{align}
where the {\it observed} nonlinearity is 
\begin{align}
\beta_{\rm NL}^{\rm obs}=\rho(T_0)\beta_{\rm NL}^{\rm cw}
\end{align}
and  the correction factor \textit{only} depends on the pump pulse duration, see Fig. \ref{fig:rho_numerical}. 
Eq. (\ref{eq:NLSE-int-long-pulse-limit-rho-main}) can be solved analytically:
\begin{align}
\label{eq:TPA-pulsed}
P(L,\tau)=P(0,\tau)e^{-\alpha L}/[1+\beta_{\rm NL}^{\rm obs} P(0,\tau) L_{\rm eff}]
\end{align}
where $L_{\rm eff}=(1-e^{-\alpha L})/\alpha$ is the effective waveguide length. For a cw field with input power $P_0$ this solution is simply $ P(L)=P_0e^{-L\alpha}/(1+\beta_{\rm NL}^{\rm obs} P_0L_{\rm eff})$, which was the form used in \cite{Lysenko2016b}. However, for a pulsed field it is not as trivial as that, because we need to calculate the average power $\bar P(L)$ by integrating the pulse train over time. For a Gaussian pulse $P(0,\tau)=P_0e^{-\tau^2/T_0^2}$ the average power is $\bar P(0)=f_{\rm rep}\sqrt{\pi}T_0P_0$, where $P_0$ is the peak power and $f_{\rm rep}$ is the repetition rate of the laser pulse train. We mention here that if the pulse is chirped, $T_0$ relates to the duration of the chirped pulse but not the transform-limited duration. Calculating the average power from Eq. (\ref{eq:TPA-pulsed}) gives
\begin{align}
\label{eq:TPA-avg}
\frac{\bar P(L)}{\bar P(0)e^{-\alpha L}}&=-\frac{{\rm Li}_{\frac{1}{2}}(-P_0\beta_{\rm NL}^{\rm obs} L_{\rm eff})}{P_0\beta_{\rm NL}^{\rm obs} L_{\rm eff}}
\\&\label{eq:TPA-avg1}
=-\frac{{\rm Li}_{\frac{1}{2}}(-\bar P(0)\beta_{\rm NL}^{\rm obs} L_{\rm eff}/[f_{\rm rep}\sqrt{\pi}T_0])}{\bar P(0)\beta_{\rm NL}^{\rm obs} L_{\rm eff}/[f_{\rm rep}\sqrt{\pi}T_0]}
\end{align}
where ${\rm Li}_{s}(x)=\sum_{k=1}^\infty k^{-s} x^k$ is the polylogarithmic function of order $s$. Note that for a sech-shaped input pulse $P(0,\tau)=P_0 {\rm sech}^2(\tau/T_0)$ this expression becomes
$\frac{\bar P(L)}{\bar P(0)e^{-\alpha L}}=\frac{{\rm atanh}\left(\sqrt{P_0\beta_{\rm NL}^{\rm obs} L_{\rm eff}/(1+P_0\beta_{\rm NL}^{\rm obs} L_{\rm eff}) }\right)}{
\sqrt{P_0\beta_{\rm NL}^{\rm obs} L_{\rm eff}(1+P_0\beta_{\rm NL}^{\rm obs} L_{\rm eff})}}$, which is almost identical in shape as the polylogarithmic function stated above. 
Converting to the average input power before the waveguide $\bar P_{\rm in}=\bar P(0)/\sqrt{C}$ and the output power after the waveguide $\bar P_{\rm out}=\bar P(L)
\sqrt{C}$, where $C$ is the total coupling loss for the two end facets, and expanding gives 
\begin{align}
\label{eq:TPA-avg-ratio}
\frac{\bar P_{\rm out}}{\bar P_{\rm in}e^{-\alpha L}}&=-C\sum_{k=1}^\infty k^{-\frac{1}{2}}\left(-\bar P_{\rm in} \frac{\sqrt{C}\beta_{\rm NL}^{\rm obs} L_{\rm eff}}{\sqrt{\pi}T_0f_{\rm rep}} \right)^{k-1}
\\
&\simeq C\left(1-\bar P_{\rm in} \frac{\sqrt{C}\beta_{\rm NL}^{\rm obs} L_{\rm eff}}{\sqrt{2\pi }T_0f_{\rm rep}}\right)
\label{eq:TPA-avg-approx}
\end{align}
The latter approximation holds for weak peak powers, so that $P_0\beta_{\rm NL}^{\rm obs} L_{\rm eff}\ll 1$. This was the approximation used in \cite{Lysenko2016,Lysenko2016a}\footnote{Note, however, that while Eq. (24) in \cite{Lysenko2016a} is correct, Eq. (25) is wrong and should be replaced by Eq. (\ref{eq:TPA-avg-approx}). Additionally, \cite{Lysenko2016} uses the relation $\beta_{\rm NL}^{\rm obs} \simeq \frac{b}{a} f_{\rm rep}T_0\alpha$, which should be corrected by a factor to read $\beta_{\rm NL}^{\rm obs} \simeq \frac{b}{a} \sqrt{2\pi} f_{\rm rep}T_0\alpha$. Finally, the fit in Fig. 3 there was carried out not as stated on the power levels inside the waveguides, but rather on the measured input and output powers outside, which means that the coupling loss affects the found fitting parameter. As a consequence the $\beta_{\rm NL}^{\rm obs}$ values and ultimately the ${\rm Im}[\chi^{(3)}]$ values should be larger by a factor $\sqrt{2\pi/C}$.}. However, as soon as the loss-normalized average power ${\bar P(L)}[{\bar P(0)e^{-\alpha L}}]^{-1}$ deviates much from unity, higher-order terms, or preferably the full polylogarithmic function Eq. (\ref{eq:TPA-avg}), should be used. We also note that $\frac{\bar P(L)}{\bar P(0)e^{-\alpha L}}\rightarrow 1$ in the linear regime. 


\section{Improved analysis of experimental data}

We performed numerical simulations of the NLSE (\ref{eq:NLSE_Au_TTM}) using a pseudo-spectral method in the interaction picture \cite{laegsgaard:2007}, where the dispersion and waveguide parameters were calculated using the accurate vectorial approach as outlined in \ref{sec:vector}, and the nonlinear coefficients were set to the experimental values from Table \ref{tab:data}. The simulations using experimentally determined nonlinearities showed that the 200 fs experiments used peak powers that were too high for restricting to the linear expansion (\ref{eq:TPA-avg-approx}) only while determining the $\beta_{\rm NL}^{\rm obs}$ value in a fit. We therefore decided to fit to the full polylogarithmic function (implemented with Matlab 2016b Nonlinear Curve Fitting Tool), and the results of a numerical sweep of the pulse durations are summarized in Fig. \ref{fig:rho_numerical}. Comparing with the semi-analytical calculations, cf. Eq. (\ref{eq:rho_semianalytical}), the full simulations show a slightly increased correction factor throughout the entire range from 30 fs to 30 ps. The main approximations behind the semi-analytical calculation are that we neglect dispersion, the temporal offset in the convolution as well as self-steepening. All these factors become important for short femtosecond pulses, where the deviation becomes most evident. Thus, the semi-analytical calculations turned out to be more accurate than expected. At the same time, the full simulation shows that the approach of using the integrated power to determine the imaginary part of the linear and nonlinear susceptibilities works extremely well, even when the nonlinearity is of a delayed nature. We found in the simulations that it was important to sample peak powers large enough to get below the 0.99 level of the normalized average power in order for the polylogarithmic fit to converge. Conversely, too large peak powers give significant nonlinear change to the pulse peak power from self-phase modulation (SPM, real part of the nonlinearity), which impedes the idea of an isolated analysis of the nonlinear loss (that indeed assumes that the pulse profile can be characterized by its input peak power). Therefore it is important that a suitable compromise is found in the power sweep. 

\begin{figure}[bt]
\centering
\includegraphics[width=\linewidth]{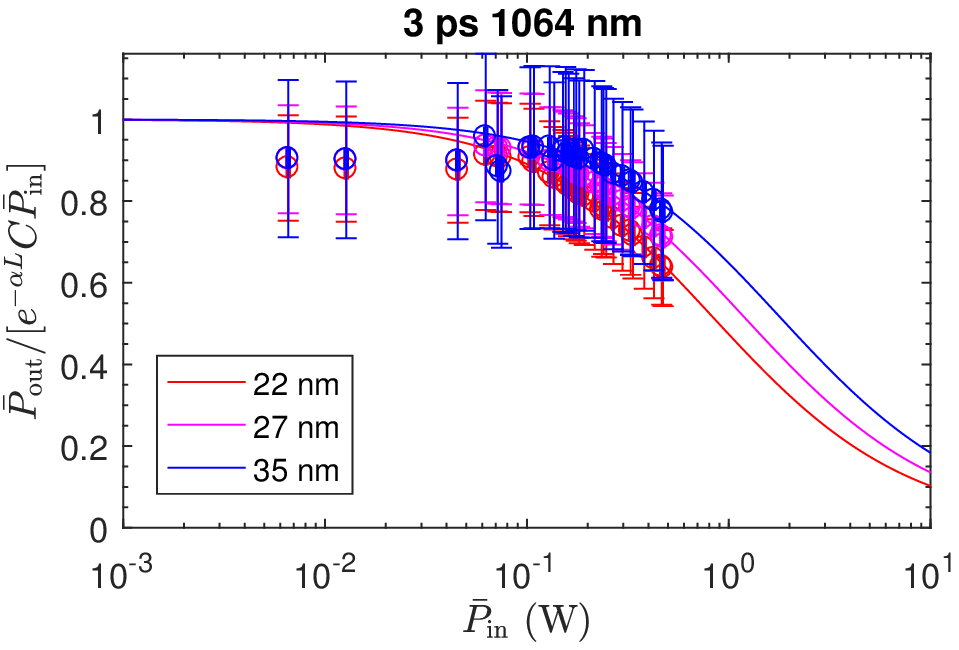}
\includegraphics[width=\linewidth]{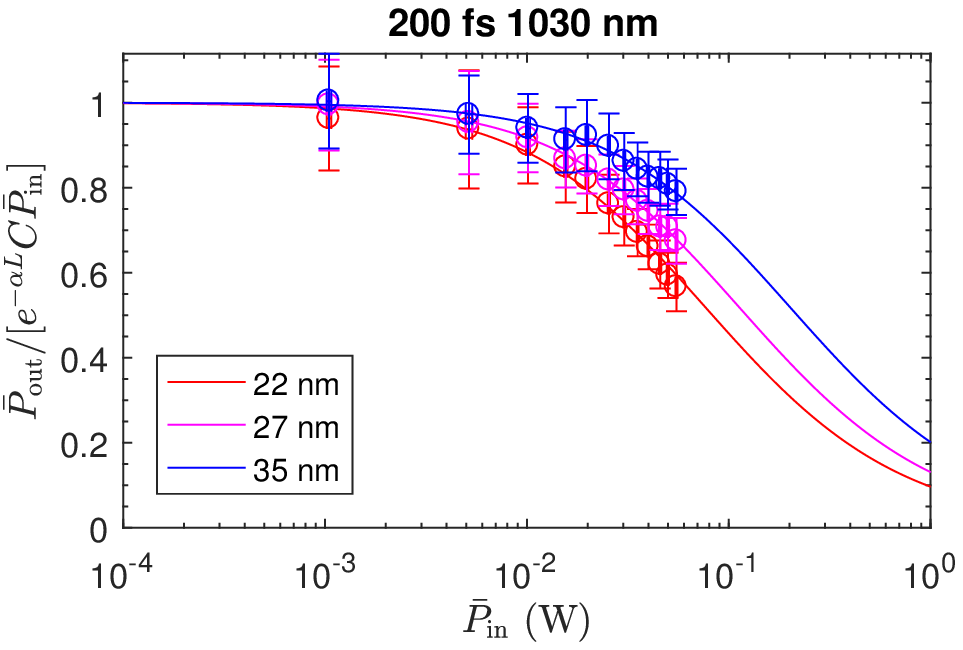}
\caption{The experimental data from \cite{Lysenko2016,Lysenko2016b} plotted vs average input power. The data shown is the average output power normalized to the average input power subtracted for coupling and propagation losses. The plot shows also the improved fit with the polylogarithmic functionality (\ref{eq:TPA-avg-ratio}), which is the exact solution of the nonlinear power loss propagation equation (\ref{eq:NLSE-int-long-pulse-limit-rho-main}). The data in (a) used 3 ps FWHM Gaussian pulses from a 1064 nm laser amplifier having $f_{\rm rep}=78$ MHz and (b) 200 fs FWHM Gaussian pulses from a 1030 nm laser amplifier having $f_{\rm rep}=200$ kHz. The average input and output powers $\bar P_{\rm in}$ and $\bar P_{\rm out}$ are those measured outside the waveguide, before correcting for the end-facet coupling losses. The gold waveguides were (a) 3.0 mm and (b)  2.0 mm long, and had 22, 27 and 35 nm gold thickness. In the 3 ps fits we excluded the three lowest average input power points as these significantly deviated from the rest, most likely due to a change in coupling loss for these measurements. 
}
\label{fig:beta}
\end{figure}

We then used the polylogarithmic fitting procedure to re-evaluate the experimental data from \cite{Lysenko2016,Lysenko2016b}. We remind that the data were recorded by using the built-in integrating function of the optical spectrum analyzer, and sweeping the power of the pump lasers in each case. "Cut-back" loss measurements on sets of nominally identical waveguide of  different lengths were used to determine the linear propagation loss $\alpha$ and the total coupling loss. 
Unlike \cite{Lysenko2016b}, to fit the data we used here two free fitting parameters $C$ and $b$, ${\bar P_{\rm out} }[{\bar P_{\rm in}e^{-\alpha L}}]^{-1}=-C {\rm Li}_{\frac{1}{2}}(-\bar P_{\rm in}\sqrt{C} b)/(\bar P_{\rm in}\sqrt{C} b)$, where $C$ represents the total coupling loss ($\sqrt{C}$ per facet). The advantage of adding the second fitting parameter $C$ in this way, instead of using the experimentally determined value, is that it automatically takes into account possible variations in the coupling loss, e.g. due to small changes in the setup. This implicitly changes the nonlinear parameter of the fit $b$ and thus the extracted $\beta_{\rm NL}^{\rm obs}=b\sqrt{\pi}T_0f_{\rm rep}/L_{\rm eff}$ value. The experimental data are shown in Fig. \ref{fig:beta} on a log-linear plot, and show excellent agreement with the fits. For convenience we shown there the average normalized output power corrected also for the fitted coupling loss parameter $C$. In this way all plots have unity as the zero power asymptote. This presentation is convenient because it immediately evidences that not only the 200 fs but also the 3 ps data show a significant departure from unity, which suggests that the first-order expansion Eq. (\ref{eq:TPA-avg-approx}) for fitting that we used in \cite{Lysenko2016} is not accurate enough. However, it also suggests that the peak power levels used in the 200 fs case were probably too high to give an accurate determination of $\beta_{\rm NL}$ since significant SPM can be expected. 

\begin{table*}
\begin{center}
\begin{tabular}{l|ccc|ccc}
& \multicolumn{3}{|c|}{200 fs 1030 nm, 200 kHz}& \multicolumn{3}{|c}{3 ps 1064 nm, 78 MHz}\\
\hline 
$t_{\rm Au}$ (nm) & 22 &  27  & 35 & 
22 &  27  & 35 \\
\hline
$\theta_{\rm Au}$ ($10^{-5}$)& 
0.303 & 0.650 & 1.67 &
0.237& 0.5143& 1.134 \\
$A_{\rm eff} ~(\mu \rm m^2)$ &
7.85 & 7.15 & 6.36 &
8.48 & 7.72 & 6.86 \\
$\tilde n$ &
1.4565 & 1.4571 & 1.4580 &
1.4557 & 1.4562 & 1.4570 \\
\hline
$C$ (dB) &    
$6.8\pm 0.1$ &   $6.7\pm 0.1$ &   $6.9\pm 0.1$ &
$2.8\pm 0.1$ &   $2.6\pm 0.1$ &   $10.5\pm 5^*$ \\
$b~(\rm W^{-1})$ &  
$42\pm5$ &  $28\pm 1$ &  $16\pm 2$ &
$2.4\pm 0.1$ &  $1.7\pm 0.1$ &  $2.7\pm 0.3$\\
$\beta_{\rm NL}^{\rm obs}~(\rm [kW\cdot m]^{-1})$ &    
$3.6\pm0.5$  &   $3.2 \pm 0.2$ &   $2.3 \pm 0.3$ & 
$1090\pm90$  &   $1060 \pm 80$ &   $2200 \pm 1500$ \\
$\beta_{\rm NL}^{\rm cw}~(\rm [kW\cdot m]^{-1})$ &    
$114\pm 15$  &   $102 \pm 7$ &   $73 \pm 9$ & 
$1480\pm 120$  &   $1440 \pm 110$ &   $2900 \pm 2000$ \\
${\rm Im}[\chi^{(3)}_{\rm obs}]~(10^{-17} {\rm m^2/V^2})$ &  
$0.58\pm 0.16$  & $0.22\pm 0.06$ &   $0.054\pm 0.015$ &
$250\pm 70$  & $100\pm 30$ &   $80\pm 60$ \\
${\rm Im}[\chi^{(3)}_{\rm cw}]~(10^{-17} {\rm m^2/V^2})$ &  
$18\pm 5$  & $6.9\pm 1.8$ &   $1.7\pm 0.5$ &
$340\pm 90$  & $140\pm 40$ &   $110\pm 80$ \\
\hline 
SBF & - & - & - & $1.27\pm 0.10$ & $1.21\pm 0.10$ & $1.24\pm 0.10$ \\
$P_0$ (kW)$^\dagger$ & - & - & - & $1.25\pm 0.02$ & $1.24\pm 0.02$ & $1.21\pm 0.02$ \\
$\gamma_{\rm NL}^{\rm obs}~(\rm [W\cdot m]^{-1})$ & - & - & - & 
$3.4\pm 0.5$ & $4.1\pm 0.5$ & $6.5\pm 4.3$ \\
$\gamma_{\rm NL}^{\rm cw}~(\rm [W\cdot m]^{-1})$ & - & - & - & 
$4.6\pm 0.6$ & $5.5\pm 0.7$ & $8.5\pm 5.8$ \\
${\rm Re}[\chi^{(3)}_{\rm obs}]~(10^{-15} {\rm m^2/V^2})$ & - & - & - & 
$15\pm 4$ & $7.7\pm 2.2$ & $4.8\pm 3.5$ \\
${\rm Re}[\chi^{(3)}_{\rm cw}]~(10^{-15} {\rm m^2/V^2})$ & - & - & - & 
$21\pm 5.9$ & $10\pm 2.9$ & $6.5\pm 4.7$ \\
\end{tabular}
\end{center}
\caption{Overview of the improved analysis of the experimental data. The top section shows the waveguide parameters as calculated using the vectorial analysis, cf. \ref{sec:vector}, the middle section - the nonlinear loss analysis, and the bottom section - the spectral broadening analysis. The correction factors used to calculate the cw nonlinear coefficients were $\rho=0.0318$ (200 fs) and $\rho=0.7376$ (3 ps). To calculate $L_{\rm eff}$ we used the measured propagation loss, cf. Table 1 in \cite{Lysenko2016}. All the conversions from waveguide nonlinearities to nonlinear susceptibilities are assigned with a 25\% error as to take into account various uncertainties in calculating the waveguide parameters. 
\\$^*$This error was estimated because the coupling loss from the fit was 7 dB above the estimated coupling loss from the low-power insertion loss analysis on 2, 3 and 4 mm waveguides that was used to also assert the propagation loss, cf. Table 1 in \cite{Lysenko2016}. 
\\$^\dagger$Note that the power level reported in \cite[Fig. 5]{Lysenko2016} was measured before the waveguide. We could not confirm that the measurements of the spectral broadening should have the high coupling loss at 35 nm as observed for the nonlinear loss measurements, so we kept the coupling loss at the level of the low-power insertion loss measurements, i.e. $C=$ 3.5 dB per two facets, which was used to calculate the estimated peak power.}
\label{tab:data}
\end{table*}

The next step is to use the fits to calculate the nonlinear susceptibilities. The found fitting parameters $b$ were then used to find the $\beta_{\rm NL}^{\rm obs} $ values. We stress that this corresponds to the \textit{observed} nonlinearity at that given wavelength and pulse duration. The values corresponding to the cw limit can then be found by applying the numerical correction factors. The data are summarized in Table \ref{tab:data}. 
The updated waveguide nonlinearities can then be used to calculate the ${\rm Im}[\chi_{\rm obs}^{(3)}]$ values, which requires calculating the waveguide parameters. 

In \cite{Lysenko2016} the real part of the waveguide nonlinearity, related to SPM, was estimated by studying the spectral broadening factor (SBF), i.e. the ratio of the bandwidth of the output pulses to that of the input pulses. This is a very rough estimate as the SBF is only known analytically for special cases, and for an unchirped pulse the well-known expression is ${\rm SBF}\simeq \sqrt{2 e^{-1}} \gamma_{\rm NL}^{\rm obs}P_0 L_{\rm eff}\simeq 0.86 \gamma_{\rm NL}^{\rm obs}P_0 L_{\rm eff}$ \cite{agrawal:2012}. This expression was used in \cite{Lysenko2016}, and it only requires estimating the peak power $P_0$ and the effective propagation length. Here we adopt a more accurate numerical calculation of the SBF by performing the 2-nd order moment average \cite{agrawal:2012}. Additionally, it is well known that a nonzero $\beta_{\rm NL}$ limits the maximum nonlinear phase shift obtainable with SPM \cite{Yin2007}, and thus the amount of spectral broadening. Here we address this issue by relating the calculated SBF to $\gamma_{\rm NL}^{\rm obs}$ by using a more accurate expression that holds for an unchirped Gaussian pulse in presence of a nonzero $\beta_{\rm NL}^{\rm obs}$:
\begin{align}
\label{eq:SBF}
{\rm SBF}\simeq \gamma_{\rm NL}^{\rm obs} P_0 L_{\rm eff} 0.86 (1+\beta_{\rm NL}^{\rm obs}P_0 L_{\rm eff} e^{-\tfrac{1}{2}})^{-1}
\end{align}
Since we have already independently determined $\beta_{\rm NL}^{\rm obs}$, we can use this to calculate a more accurate value of $\gamma_{\rm NL}^{\rm obs} $ for a given SBF. Eq. (\ref{eq:SBF}) was calculated in the usual manner by maximizing the  chirp across the temporal envelope and relating that to the SBF. The equation is only approximate since the temporal position of maximum chirp used in the calculation was taken $\tau_{\rm max}/T_0=1/\sqrt{2}$, just as in the $\beta_{\rm NL}^{\rm cw}\rightarrow 0$ case, but we checked that this is an excellent approximation. We also note that the $\beta_{\rm NL}^{\rm obs}P_0 L_{\rm eff}$ levels were around 0.5-1.0 in the 3 ps case. The updated values for the SPM analysis are also listed in Table \ref{tab:data}; only data for the 3 ps experiment are presented as the SBF was not investigated for the 200 fs case. 

The conversion from the waveguide nonlinearity to the real part of the nonlinear susceptibility of gold included detracting the estimated contributions from the cladding materials to the SPM, i.e. using the connection in Eq. (\ref{eq:gamma-beta-cw}), the updated field fractions shown in \ref{sec:vector} and the nonlinearities of the cladding materials as in \cite{Lysenko2016a}. We do note that since gold exhibits the dominating nonlinearity, the effect of including the cladding nonlinearity in this calculation was small. 
We also mention that the peak powers were found by assuming a 3 ps Gaussian pulse, which may present a significant uncertainty in the analysis since we know the pump spectrum supports sub-ps pulses, and that the pump pulse is therefore chirped. While this does not affect the estimate of the peak power, it is well known that the spectral broadening dynamics is significantly different when the pulse is chirped. Indeed, calculating the SBF for a chirped pulse is impossible analytically, and it is therefore useful to implicitly verify the calculated parameters by performing full simulations of the NLSE. This analysis follows next. 

Before that, let us comment on the updated experimental values. Firstly, all the data still confirm the $1/t_{\rm Au}$ trend, i.e. that the bulk thermomodulated nonlinearity increases for thinner waveguides. 
Secondly, let us look at the data for the imaginary part in the cw limit. There we have data for both pulse durations, and these data have been corrected to reflect that the 200 fs and 3 ps experiments have different correction factors in the delayed nonlinearity of gold. It is clear that they do not match each other: there is an order of magnitude difference between them. Interestingly, though, the ratio between the 22 and 27 nm cases is the same, close to 20, while the 35 nm case is not showing this (probably due to the large coupling losses). This constant ratio might be related to the less ideal circumstances in the 3 ps case, and certainly warrants new experiments with more ideal pump pulses. It could also be because the 200 fs data are underestimated. Finally, we remark that the cw nonlinear susceptibilities are 1-2 orders of magnitude larger than the near-IR values predicted by Marini et al. \cite{Marini2013}, i.e. Eq. (\ref{eq:chi3_T}). 

\section{Numerical simulations of the NLSE}

\begin{figure}[bt]
\centering
\includegraphics[width=\linewidth]{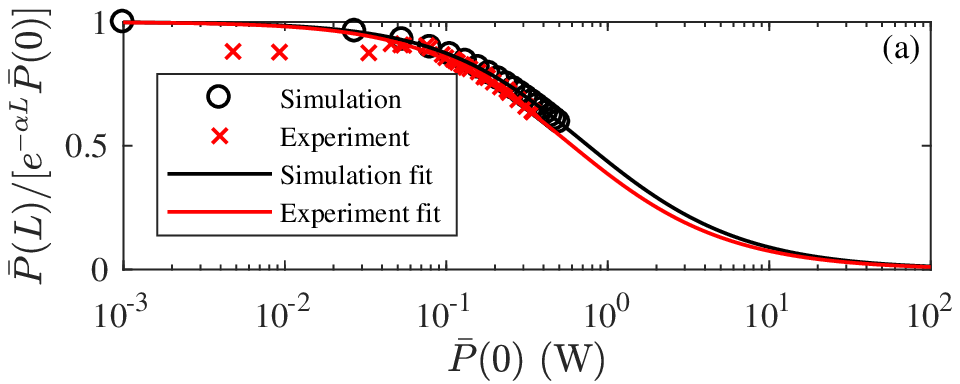}
\includegraphics[width=\linewidth]{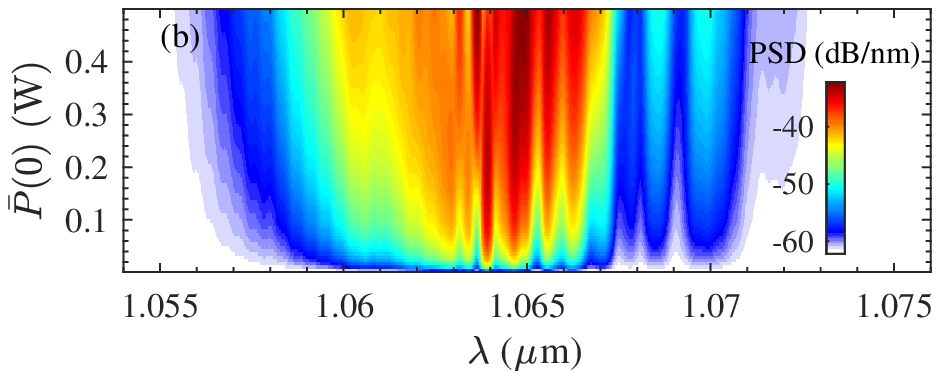}
\caption{Numerical power sweep from 0.1-500 mW average power inside the waveguide for the 22 nm waveguide case (3 mm long) pumped at $\lambda=1.064\mic$ with 3 ps 78 MHz pulses (up to 1.5 kW peak power). (a) Normalized power loss vs. input power, compared directly to the experimental data. (b) The spectral evolution during the power sweep. The simulation used $\beta_{\rm NL}^{\rm cw}$ and $\gamma_{\rm NL}^{\rm cw}$  from Table \ref{tab:data}. The simulation took the experimentally measured spectrum as input, and the quadratic phase across the spectrum was adjusted to give a 3 ps initial pulse duration (group-delay dispersion ${\rm GDD}=5.1\cdot 10^5~\rm fs^2$). 
}
\label{fig:loop_3ps}
\end{figure}

We now show full simulations of the NLSE to verify the 3 ps case. Generally the numerical simulations expose excellent agreement concerning the loss evolution  through linear and nonlinear effects: for given values of $\alpha$ and $\beta_{\rm NL}^{\rm cw} $ the loss calculated numerically could be directly compared to that of the experiments. This verifies the notion that measuring the loss by observing the integrated power alone can be used to reconstruct the imaginary part of the nonlinearity. As a representative case, Fig. \ref{fig:loop_3ps} shows the results of a numerical power sweep for the 3 ps case, where as input we used the measured  spectrum, which was quite modulated during the amplification stage, as well as the $\beta_{\rm NL}^{\rm cw}$ and $\gamma_{\rm NL}^{\rm cw} $ from Table \ref{tab:data} for a 22 nm waveguide. The nonlinear loss characterization is in very good agreement with the experimental data. The main uncertainty in the 3 ps case is the chirp across the input pulse: the experimental spectrum supports sub-ps pulse durations, and in the simulations we imposed a linear positive chirp to stretch the pulse to 3 ps. It is well known that chirped pulses experience completely different nonlinear effects concerning SPM (just flipping the chirp sign can change spectral broadening to spectral compression), but the impact on nonlinear loss is not studied in details. We did confirm that we saw almost identical simulation results with a negative chirp, and we even found that a transform-limited Gaussian 3 ps pump pulse with the same pulse energies showed the same trends with respect to nonlinear loss. Thus, it seems that the chirp is not so critical, and the main issue is the actual pulse duration (and not whether it is transform limited or not). We will not pursue this further here, but this is certainly worth to investigate in a future publication. 

For this simulation we also show the spectral evolution, which can be used to get an impression of SPM from the real part of the nonlinear susceptibility. On a qualitative level we could observe spectral broadening similar to that of the experiment: a blue shoulder below 1062 nm similar to that of Fig. 5(b) in \cite{Lysenko2016} was also seen in the simulations. On a quantitative level it seemed more prominent using higher values of $\gamma_{\rm NL}^{\rm cw}$. We can therefore tentatively conclude that it seems that the waveguide SPM nonlinearity as calculated from the SBF analysis lies in the correct range, but there are indications that it might be a factor 2-3 higher. A more precise determination would require experiments with transform-limited pulses, where modulations in the output pulse spectrum can be identified as stemming from SPM. 

\begin{figure}[bt]
\centering
\includegraphics[width=\linewidth]{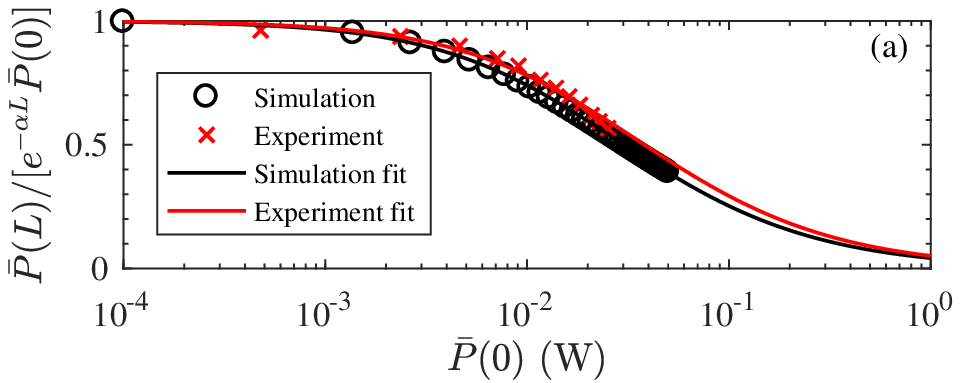}
\includegraphics[width=\linewidth]{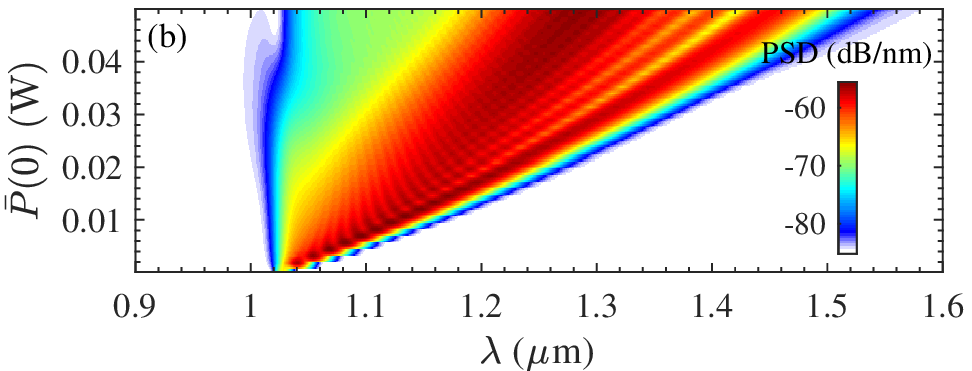}
\caption{Numerical power sweep from 0.01-50 mW average power inside the waveguide for the 22 nm waveguide case (2 mm long) pumped at $\lambda=1.03\mic$ with 200 fs 200 kHz pulses (up to 800 kW peak power). (a) Normalized power loss vs. input power, compared directly to the experimental data. (b) The spectral evolution during the power sweep. The simulation used $\beta_{\rm NL}^{\rm cw}$ for the 200 fs case and $\gamma_{\rm NL}^{\rm cw}$ for the 3 ps case from Table \ref{tab:data}. 
}
\label{fig:loop_200fs}
\end{figure}

In Fig. \ref{fig:loop_200fs} we show a similar simulation performed for the 1030 nm case, using an unchirped 200 fs Gaussian pump pulse. The nonlinear real part of gold was kept the same, but the imaginary part was adjusted to the experimentally determined value, cf. Table \ref{tab:data}. Again, when comparing with the experimental data the simulations show excellent agreement regarding the nonlinear power loss analysis. 
The lower repetition rate means that the 50 mW average power outside the waveguide corresponds to close to 1 MW peak power inside the waveguide. This is over 2 orders of magnitude more than in the 3 ps case, and in this case there is clear spectral broadening towards the red even for low average powers. The red-shift is caused by the delayed nonlinearity of gold (also observed in simulations at 800 nm \cite{Marini2013}). 
Using 200 fs pump pulses means that the correction factor is small, making the nonlinear response of gold less dominating. We noted that the cladding nonlinear effects start to become significant in this simulation, but also verified that the main nonlinear dynamics remained the same even when the cladding nonlinear coefficients were set to zero. 

It is interesting to look towards longer wavelengths, where linear losses are much lower and the gold nonlinearities stay more or less the same. In particular, at 1550 nm the 10 nm thick state-of-the-art gold waveguides with up to 7 mm lengths exhibit propagation losses as low as 0.62 dB/mm \cite{kinsey:2015}. The promising feature of such low losses is that they allow for cm-scale waveguides. Inspired by these numbers, we performed  simulations at 1550 nm. For simplicity we kept the same waveguide thickness (22 nm) as in the previous simulations, and used both the measured linear loss (our waveguides were characterized at 1550 nm to have a 2.6 dB/mm loss) and a  lower linear loss (0.5 dB/mm\footnote{Note that the low loss of 0.62 dB/mm obtained by \cite{kinsey:2015} is in part due to the longer wavelength, but in part also caused by the thin gold layer, which favors lower loss. However, as we outline below, the thin waveguides suffer from a low nonlinear field overlap in gold.}). 
At 1550 nm, $\theta_{\rm Au}\simeq 1.5\cdot 10^{-7}$ and $A_{\rm eff}\simeq 21 ~\rm \mu m^2$ for our waveguide design, cf. Fig. \ref{fig:dispersion}. The low power fraction in gold gives 1-2 orders of magnitude lower effective nonlinear susceptibility than at 1030 nm, additionally the larger mode area also decreases nonlinear effects. Our NLSE simulations show that for 100 fs pump pulses (e.g. using an ultrafast Er fiber amplifier), the nonlinear effect that can be observed (weak spectral broadening) is mainly due to the cladding materials. Gold is not dominating here because of the reduced contribution to the waveguide nonlinearity as outlined above and due to the significant reduction of the delayed gold nonlinear response for a 100 fs pulse due to the correction factor on the order of  $\rho \simeq 0.01$. 
If we take an even thinner waveguide, e.g. 10 nm, the field fraction in gold is even smaller, but at the same time the $1/t_{\rm Au}$ enhancement might give some of this lost terrain back. This might be interesting to investigate further. 

Finally, we also investigated the role of linear and nonlinear losses. In the simulations of the experiments presented in this section the linear loss accounts for 20-30 dB, and the nonlinear loss for additional 5-10 dB. Combined with 6-10 dB coupling loss this means that at the exit of the waveguide we have 40-50 dB power loss. This is a significant challenge, not only because very little light comes out of the waveguide (which effectively hindered us from measuring the spectra in the 200 fs case due to the low repetition rate) but also because very high peak powers are needed to see nonlinear effects, typically kW for picosecond pulses and sub-MW for femtosecond pulses. We therefore attempted to carry out simulations with zero linear loss, hoping to alleviate this power-hungry scenario, but generally the main obstacle lies in the quite large imaginary part of the nonlinearity. This significantly limits the amount of spectral broadening and is an obstacle for using nonlinear LRSPP waveguides for low-power purposes. 
A similar problem was historically encountered when silicon was promoted as a new promising nonlinear material due to its extremely large nonlinearities, 2-3 orders of magnitude larger than glass. However, the presence of a substantial imaginary nonlinear susceptibility gave rise to significant two-photon absorption, effectively limiting continuum generation to bandwidths well below an octave \cite{Koonath2008}.

\section{Discussion and Conclusion}

This paper provides a review of recent theory and experimental work in our lab \cite{Lysenko2016,Lysenko2016a,Lysenko2016b} concerning long-range surface plasmon polaritons propagating in thin gold strip waveguides sandwiched with symmetric layers of dielectric cladding materials. In contrast to most studies in gold, we here focus on the IR range where the linear and nonlinear losses are lower. 

We have thoroughly updated the analysis of the experimental data recorded and published in \cite{Lysenko2016,Lysenko2016b} for 200 fs and 3 ps $1\mic$ pulsed lasers. The updated data for the material (bulk) nonlinear susceptibilities of gold show values 1-2 order of magnitude higher than the theoretical values predicted in Eq. (\ref{eq:chi3_T}). These values are also higher than those previously reported by us \cite{Lysenko2016,Lysenko2016b}, which in part is due to an updated theory, an improved analysis (better fitting routines and more accurate formulas) and fixing of various issues with the experimental data. The data showed an increased nonlinear response as the gold thickness is reduced, scaling similarly to the phenomenological $1/t_{\rm Au}$ absorption dependence, thus confirming what was stated in this direction in \cite{Lysenko2016,Lysenko2016b}. 

A significant effort was devoted to updating the theory for the strip waveguides published in \cite{Lysenko2016a} by taking into account the vectorial nature of the modes. This update makes the nonlinear Schr\"odinger equation (NLSE) \cite{Lysenko2016a,Lysenko2016b} consistent with previous literature, especially the seminal theory by Marini et al. \cite{Marini2013}. It also updates some erroneous formulas that unfortunately appeared in \cite{Lysenko2016a}. 

The simulations of the full NLSE, including the nonlinear effects from the delayed hot-electron dynamics and the cladding dielectric materials, clearly show that the experimental protocol we used to measure the imaginary part of the nonlinear susceptibility is very accurate. The only caveat is that for large peak powers significant self-phase modulation may occur through the real part of the waveguide nonlinearity, which is not ideal for an isolated measurement of the imaginary nonlinearity. 

The improved waveguide mode calculations show that in the IR the 20-35 nm thin gold strip waveguides have nonlinear field fraction values on the order of $10^{-7}$-$10^{-5}$. Thus, despite the large $\chi^{(3)}$ values of metals, that are typically 5-8 orders of magnitude larger than typical values for dielectric materials, the effective nonlinear contribution of gold is not necessarily larger than that of the cladding materials, where the nonlinear field fractions are close to unity. 

Our analysis and simulations of the NLSE shows that for the conditions of the  experiments (1064 nm and 3 ps pulse duration, 1030 nm and 200 fs pulse duration), gold has the dominating contribution to the measured nonlinearity of the waveguides. This allows us to study the dependence of the pulse duration on the measured imaginary part of the nonlinear susceptibility of gold. The main nonlinear response from gold is due to self-action of the pump pulse, where hot electrons are created that gives a cubic nonlinear response. The delayed nature of the hot electron dynamics gives a purely non-instantaneous response, and as we pointed out recently \cite{Lysenko2016b}, it significantly inhibits nonlinearity when femtosecond instead of picosecond pulses are involved. By performing simulations of the NLSE using pulse durations in the range 30 fs to 30 ps, we here provide a confirmation of this correction factor, which  was calculated previously using a semi-analytical approach \cite{Lysenko2016b}. The effect was also confirmed by direct simulations of the experimental conditions. 
This emphasizes that one has to take into account also the pulse duration when comparing literature measurements of the nonlinearity of metals.

While there is no doubt that the correction factor is needed to understand the nonlinear strength for a certain pulse duration, we were unable to unify the experimental data for 200 fs and 3 ps. Using the waveguide measurements of the nonlinear loss parameter, we can calculate the bulk nonlinear susceptibility associated with this response, and then adjust for the pulse duration dependence through the correction factor. This should give the same values for the 200 fs and 3 ps experiments, but there is still a factor 20 missing that we cannot account for at the present moment. This remains to be resolved in future work. 


Clearly, when compared to 1030 nm, 1550 nm is a better wavelength for observing propagation-related nonlinear effects for surface plasmon polaritons, like self-phase modulation, simply because the linear loss is much lower. However, the challenge is that the nonlinear response also drops orders of magnitude, making the effective waveguide nonlinearity of gold comparable to that of the cladding materials. This issue might be resolved with some more advanced waveguide designs. 

One promise of metal waveguides is that the huge metal nonlinearities promote nonlinear effects with low peak powers. However, our analysis indicates that 10s or 100s of kW of peak powers are needed to see effects like the supercontinuum generation. This is simply because the nonlinear field fraction in the metal is extremely low and because the nonlinearity is purely delayed, which especially penalizes femtosecond pumping. Additionally, the quite large imaginary nonlinearity of gold, leading to nonlinear losses, also prevents significant nonlinear interaction with low peak-power pulses. 

Our simulations and analysis in the near-IR are based on the fact that the nonlinear response of gold is determined by the thermo-modulational effect (self-excitation of hot electrons). This effect dominates in the short-wavelength near-IR, while the ponderomotive nonlinearity, cf. Eq. (\ref{eq:chi3_PM}), in the mid-IR becomes comparable or larger than the hot-electron contribution due to a favorable $1/\omega^4$ scaling. 
This points towards pumping further into the IR, where we should start seeing significant contributions from the ponderomotive nonlinearity. 
Additionally, since the ponderomotive nonlinearity is instantaneous it does not suffer from having a reduced nonlinear response for ultrashort pulses. Therefore we expect that for femtosecond pump pulses the ponderomotive nonlinearity will dominate the thermo-modulational nonlinearity beyond $\lambda>2000$ nm. 

Unfortunately the strip waveguides we used in our numerical model were designed mostly for the 1000 nm range, and as we see already at 1550 nm the fraction of light in gold is $\sim 10^{-7}$. Therefore we cannot at this time assess the potential in the near- to mid-IR range. We therefore propose to design new waveguides specifically for this range, with decent power fractions $10^{-5}$ in gold and where the ponderomotive nonlinearity can dominate for ultrashort pump pulses. 
Another exciting prospect of longer wavelengths is of course that one may exploit pumping in the anomalous dispersion regime. This will allow for temporal soliton formation leading to very significant spectral broadening and other intriguing effects, which has yet to be observed from the action of the ultrafast nonlinearity of gold.

\section{Acknowledgements} Andrea Marini, Binbin Zhou and Radu Malureanu are acknowledged for fruitful discussions. AVL acknowledges Villum Fonden for partial support. 


\appendix

\section{The nonlinear response of gold}
Following Marini et al. \cite[Eq. (17)]{Marini2013}  one can derive the thermo-modulational interband nonlinear response based on the two-temperature model:
\begin{align}\label{eq:PNL-Au}
\bar P_T^{(3)}(z,t)\simeq &
\chi_{T,\rm cw}^{(3)}(\omega_0)E(z,t)
\times
\\\nonumber&
\int_{-\infty}^\infty ds h_T(t-s)|E(z,s)|^2
\end{align}
where $\chi_{T,\rm cw}^{(3)}(\omega_0)$ is the value of the nonlinear susceptibility of gold in the cw limit, which can be calculated from the thermo-modulational interband response Eq. (\ref{eq:chi3_T}). The temporal response function and it's Fourier transform are given by \cite{Marini2013}
\begin{align}\label{eq:hT}
h_T(t)=\Theta(t)\frac{1}{\tau_{\rm th}-\tau_{\rm r}}(e^{-t/\tau_{\rm th}}-e^{-t/\tau_{\rm r}}), \\ \tilde h_T(\omega)=\frac{1}{(1-i\omega \tau_{\rm th}) (1-i\omega \tau_{\rm r})}
\end{align}
In our calculations we took $\tau_{\rm th}=333$ fs and $\tau_{\rm r}=1.04$ ps. 

The hypothesis posed in \cite{Lysenko2016,Lysenko2016a,Lysenko2016b} is that this thermo-modulational nonlinearity is enhanced as the gold film is made thinner. This is due to the well-known fact that the absorbed power increases for nanometer-scale particles, which is usually modeled by adding a $1/R_{\rm Au}$ term on the imaginary part of the permittivity, where $R_{\rm Au}$ is the gold radius. We proposed to extend this to thin slab waveguide by considering a generalized Drude model for the metal permittivity \cite{Lysenko2016a}
\begin{align}\label{eq:Drude}
\varepsilon_m(\omega)=1-\frac{\omega_p^2}{\omega^2+i \gamma_f(t_{\rm Au})\omega}
\end{align}
Here $\omega_p$ is the plasma frequency and $\gamma_f$ is the collision frequency of the electrons. This is the parameter that is size dependent, which we model by
\begin{align}\label{eq:gamma_f}
\gamma_f(t_{\rm Au})=\gamma_{f,\infty}+\frac{v_F}{t_{\rm Au}}
\end{align}
where $\gamma_{f,\infty}$ is the bulk value of the collision frequency, $v_F$ is the Fermi velocity of gold and $t_{\rm Au}$ is the gold layer thickness. Note that the prefactor in front of the $1/t_{\rm Au}$ term is arbitrary, but historically for spheres it is taken unity \cite{kreibig1995} unless otherwise stated. It follows directly from Eq. (\ref{eq:gamma_f}) that the thermo-modulational nonlinearity is also changed. Since $\chi_{T}^{(3)}\propto {\rm Im}[\varepsilon_m]$ the nonlinearity will become
\begin{align}
\label{eq:chi3_TTM_tAu}
\chi_{T,\rm cw}^{(3)}(\omega_0,t_{\rm Au})=\chi_{T,\rm cw}^{(3)}(\omega_0)\left(
1+\frac{v_F}{\gamma_{f,\infty}t_{\rm Au} }\right)
\end{align}
In order to have consistent values for this scaling, we remark that the Fermi velocity of gold is usually taken $v_F=1.4\cdot 10^{6}$ m/s. However, the Drude model parameters $\omega_p$ and $\gamma_{f,\infty}$ are not consistent in the literature \cite{Olmon:PhysRevB.86.235147}, which in part can be ascribed to different types of gold and other properties such as temperature. We are mainly concerned about the infrared behaviour (as losses are much lower), and here it is relevant to remark that different sets of fitting parameters $(\omega_p,\gamma_{f,\infty})$ in the Drude model lead to quite similar infrared behaviour, so it is always important to use them pairwise. Our waveguides are made from sputtered gold, so we choose to model the infrared behaviour using the Drude model fit to the evaporated gold sample data from \cite{Olmon:PhysRevB.86.235147}, i.e. a collision time $\tau_{f,\infty}=1/\gamma_{f,\infty}= 14$ fs and a plasma frequency $\omega_p=8.5$ eV$=1.29\cdot 10^{16}$ rad/s. It turns out that this is almost identical to the Drude model used by, e.g., Marini et al. \cite{Marini2013}, although they use  quite different values for $\omega_p$ and $\tau_{f,\infty}$. Instead, it is a factor 2 larger than the standard Drude model we used in \cite{Lysenko2016a}, which took $\omega_p=9.1$ eV and $\tau_{f,\infty}=30$ fs. 

As we model the size-dependent enhancement by a Drude model, it is relevant to remark that it only models intraband effects, and it therefore breaks down below $\lambda=800$ nm, where the permittivity starts getting significant contributions from the interband effects. For a complete model of $\varepsilon_m$  the fit in \cite{Rakic1998} is recommended\footnote{The infrared permittivity of that fit is a factor of 2 larger than the data for evaporated gold by Olmon et al. \cite{Olmon:PhysRevB.86.235147}.} (which was used by Marini et al. in \cite[Fig. 3]{Marini2013}), but it is less clear how the size-dependent behaviour should be included in the interband effects.

In the TTM, the nonlinear response of gold is purely delayed. There is also a possible contribution to an instantaneous response through the ponderomotive force \cite{Ginzburg2010}\footnote{Note that Eq. (\ref{eq:chi3_PM}) is corrected for a typo, as \cite[Eq. (5)]{Ginzburg2010} should have used $\hbar^3$ instead of $\hbar$ in the denominator of the parenthesis.}
\begin{align}\label{eq:chi3_PM}
\chi^{(3)}_{\rm PM}(\omega_0)=\frac{3}{2} \hbar^{-2} \left(\frac{\omega_p}{3 \pi^2  \varepsilon_0 m_e e}\right)^{2/3} (e/\omega_0)^4
\end{align}
As this scales as $\omega^{-4}$ it only becomes relevant in the IR, hitting $\chi^{(3)}_{\rm PM}\simeq 10^{-18}~\rm m^2/V^2$ around 1500 nm. Instead, at 1000 nm the value is quite low, $\chi^{(3)}_{\rm PM}\simeq 10^{-19}~\rm m^2/V^2$, and therefore it is a good approximation to neglect this effect in extracting the experimental data for the real part of $\chi^{(3)}$ in \cite{Lysenko2016}, which were on the order $10^{-15}~\rm m^2/V^2$. 

\section{Nonlinear Schr\"odinger equation }
\label{sec:vector}

We here show the basic connection between the complex nonlinear susceptibility and the waveguide nonlinear parameters. This connection is established through the nonlinear Schr\"odinger equation (NLSE), which here is reported for the basic geometry considered in our recent experiments: a thin gold strip waveguide surrounded by two dielectric cladding layers (TaO$_5$ and silica). The gold layer of the strip waveguide was $t_{\rm Au}$ nm thick in the $y$ direction (samples with 22, 27 and 35 nm were used), the following layer on each side was a $t_2=26$ nm TaO$_5$ layer, followed by a silica layer. In  the other transverse direction, $x$, the gold and TaO$_5$ layers were $w=10 \mic$ wide. 

In our previous work \cite{Lysenko2016a} we solved Maxwell's equations to find the waveguide modes, and considered only the transverse electric field component in the $y$ direction. Unfortunately, the definitions of the modal contributions to the nonlinearity in each material section \cite[Eqs. (9-11)]{Lysenko2016a} are wrong; since the nonlinearity is third-order, the fields should have been taken to the $4^{\rm th}$ power instead of $2^{\rm nd}$. As a consequence the nonlinear field overlap fractions $\theta_{\rm Au}$ are two orders of magnitude  smaller. 

Here we turn to a vectorial approach based on \cite{Marini2013}, which includes the electric field component in the longitudinal $z$ direction too. Thereby we not only correct the error made in our previous publication, we also use a more accurate vectorial notation in agreement with other results in the literature \cite{Laegsgaard:2007-OE,Marini2013,kolesik-2004-PhysRevE.70.036604,afshar.2009}. 
The initial calculations are the same. Specifically, the equation for the the infinite slab \cite[Eq. (7)]{Lysenko2016a} is first solved to find the eigenvalue $\beta_\infty=n_\infty \omega_0/c$. For the considered geometry only transverse magnetic solutions are present, and we constrict ourselves to the even solutions. This is followed by the correction due to the finite slab width in the $x$ direction, and by solving \cite[Eq. (8)]{Lysenko2016a} we obtain the mode propagation constant $\beta\equiv n_{\rm eff} \omega_0/c$. It is here important to locate the value having the highest effective index $n_{\rm eff}$ as close as possible to $n_{\infty}$ to ensure that we operate with the fundamental spatial mode in the $x$ direction. The dimensionless transverse magnetic field is then $h_x(x,y)=f_x(x)g_x(y)$ with
\begin{align}
f_x(x)&=\begin{cases}
A e^{-k_{3}(\beta)x} & x>\tfrac{1}{2}w \\
B e^{k_{3}(\beta)x} & x<-\tfrac{1}{2}w \\
C \cos(k_1(\beta) x) & |x|\leq \tfrac{1}{2}w
\end{cases}
\\
g_x(y)&=\begin{cases}
A' e^{-k_{3}(\beta)y} & y>\tfrac{1}{2}(t_{\rm Au}+t_2) \\
B' e^{k_{3}(\beta)y} & y<-\tfrac{1}{2}(t_{\rm Au}+t_2) \\
C' e^{-k_{2}(\beta)y} & \tfrac{1}{2}t_{\rm Au}<y\leq \tfrac{1}{2}(t_{\rm Au}+t_2) \\
D' e^{k_{2}(\beta)y} & -\tfrac{1}{2}(t_{\rm Au}+t_2)\leq y<-\tfrac{1}{2}t_{\rm Au} \\
E' \cosh(k_1(\beta) y) & |y|\leq \tfrac{1}{2}t_{\rm Au}
\end{cases}
\end{align}
where
\begin{align}
k_j(\beta)=\sqrt{\beta^2-\varepsilon_j\omega^2/c^2}
\end{align}
and where $j=1$ is gold, $j=2$ is TaO$_5$ and $j=3$ is silica. 

The corresponding dimensionless electric fields are then found from Ampere's law, written in the form $\vec{\nabla} \times \vec{H}=\varepsilon_0\varepsilon \frac{\partial \vec{E}}{\partial t} $. By adopting the form
\begin{align}
\vec{E}(\vec{x},t)&=\sqrt{a_P}A(z,t)\vec{e}(x,y)e^{i(\beta z-\omega_0 t)} \\
\vec{H}(\vec{x},t)&=\varepsilon_0 c\sqrt{a_P}A(z,t)\vec{h}(x,y)e^{i(\beta z-\omega_0 t)} 
\end{align}
and noting that for TM solutions only $h_x$ is nonzero, we find the relations
\begin{align}
e_z(x,y)&=\frac{i}{k_0\varepsilon(x,y)}\frac{\partial h_x}{\partial y}, \quad {\rm longitudinal} \\
e_x(x,y)&=\frac{\beta}{k_0\varepsilon(x,y)}h_x(x,y), \quad {\rm transverse}
\end{align}
where $k_0=\omega_0/c$ and $\varepsilon(x,y)$ contains is the relative permittivity of the constituent materials of the waveguide. The vectorial approach allows taking into account the longitudinal electric field when calculating the nonlinear response. 

The fields are normalized so that $|A|^2$ represents the power in the propagation direction $z$ in watts, which requires 
\begin{align}
a_P&=\frac{2}{\varepsilon_0 c s_z}\\
s_z&=|\int_{dA} {\rm Re}[\vec{e}\times \vec{h}^*]\cdot \hat z|
\end{align}
where $\int_{dA}$ implies integration over the entire waveguide cross section. This normalization \textit{de facto} renders $\vec{e}$ and $\vec{h}$ dimensionless, implying $s_z$ has the unit ${\rm m}^2$, i.e. representing the area of the mode. 

The nonlinear waveguide dynamics is then found by writing the following NLSE, which uses the notation in \cite{Marini2013}
\begin{multline}\label{eq:NLSE_Au}
\left(i\partial_{\zeta}+\hat D_\tau +i\tfrac{1}{2}\alpha\right) A+\hat S (\Gamma_{\rm TaO_5}+\Gamma_{\rm SiO_2}) A|A|^2
\\
+\hat S \Gamma_{\rm Au}A\int_{-\infty}^\infty d\tau' h_T(\tau-\tau') |A(\zeta,\tau')|^2=0
\end{multline}
where $\alpha$ is the linear loss parameter, $\hat D_\tau=\sum_{n=2}^\infty i^n \beta_{n} \partial^n/\partial \tau^n$ is the dispersion operator, $\beta_n=d^n\beta (\omega_0)/d\omega^n$ are the higher-order dispersion coefficients, and $\hat S=1+i\omega_0^{-1}\partial_ \tau $ is the self-steepening operator. The NLSE is transformed to the coordinate system  $\zeta$ and $\tau$ moving with the group velocity of the $\omega_0$ wave $v_g=1/\beta_1$.
The delayed nonlinear response of gold is modelled by the convolution of $|A|^2$ with $h_T$, the dimensionless temporal response function calculated using the two-temperature model, cf. Eq. (\ref{eq:hT}). The (complex) nonlinear waveguide coefficients are (unit ${\rm W^{-1}m^{-1}}$)
\begin{align}
\label{eq:Gamma-app}
\Gamma_{j}&=\frac{3\omega_0 \theta_j \chi^{(3)}_{j}}{4\varepsilon_0 c^2 \tilde n^2 A_{\rm eff}}
, \quad {j=\rm Au,\, TaO_5, \,SiO_2}
\end{align}
In our approach, the nonlinear susceptibility we use for gold is based on Eq. (\ref{eq:chi3_TTM_tAu}), i.e. the TTM nonlinearity modified for an ultra-thin gold layer. Note that we keep the prefactor $3/4$ for historical reasons. In the dielectric cladding materials (a) we neglect Raman effects for simplicity and (b) nonlinear absorption is vanishing in the near-IR (${\rm Im}[\chi^{(3)}]=0$), making the nonlinear coefficients for the cladding materials real. Instead for gold $\Gamma _{\rm Au}$ is complex. 
In the vectorial notation, one can define an effective mode area as follows \cite{afshar.2009}
\begin{align}
\label{eq:Aeff}
A_{\rm eff}=\frac{s_z^2}{\int_{dA} ({\rm Re}[\vec{e}\times \vec{h}^*]\cdot \hat z)^2}
\end{align}
The dimensionless field overlap integrals are
\begin{subequations}
\begin{align}
\theta_{\rm Au}&=\tilde n^2\frac{4}{3}\frac{\int_{dA_{\rm Au}} |\vec{e}|^4} 
{\int_{dA} ({\rm Re}[\vec{e}\times \vec{h}^*]\cdot \hat z)^2}
\\
\theta_{\rm TaO_5}&=\tilde  n^2\frac{1}{3}\frac{\int_{dA_{\rm TaO_5}} (2|\vec{e}|^4+|\vec{e}^2|^2)} 
{\int_{dA} ({\rm Re}[\vec{e}\times \vec{h}^*]\cdot \hat z)^2}
\\
\theta_{\rm SiO_2}&=\tilde  n^2\frac{1}{3}\frac{\int_{dA_{\rm SiO_2}} (2|\vec{e}|^4+|\vec{e}^2|^2)} 
{\int_{dA} ({\rm Re}[\vec{e}\times \vec{h}^*]\cdot \hat z)^2}
\end{align}
\end{subequations}
We stress that the numerators in the $\theta$ definitions are highly material-dependent, and here we have assumed that the cladding dielectric materials are isotropic, and that gold has a dominant nonlinearity of thermo-modulational intraband origin. We also mention that in the vectorial case the definition of the  field fractions $\theta$ means that they will not add up to unity. 

In the cw limit, i.e. for pulses much longer than the gold relaxation times characteristic of the delayed response $h_T$ (sub-picosecond range), we have that $A\int_{-\infty}^\infty d\tau' h_T(\tau-\tau') |A(\zeta,\tau')|^2\simeq A|A|^2$. Writing then  the total nonlinear response in the cw limit as $\Gamma_{\rm tot} =\gamma_{\rm NL}^{\rm cw}+i\beta_{\rm NL}^{\rm cw}/2$ we find 
\begin{align}
\gamma_{\rm NL}^{\rm cw}=&\frac{3\omega_0}{4\varepsilon_0 c^2 \tilde  n^2 A_{\rm eff}} \\&\times \nonumber
[\theta_{\rm Au}{\rm Re}[\chi^{(3)}_{\rm Au}]+\theta_{\rm TaO_5}\chi^{(3)}_{\rm TaO_5}+\theta_{\rm SiO_2}\chi^{(3)}_{\rm SiO_2}]
\\
\beta_{\rm NL}^{\rm cw} =& \frac{3\omega_0}{2\varepsilon_0 c^2 \tilde n^2 A_{\rm eff}}\theta_{\rm Au}{\rm Im}[\chi^{(3)}_{\rm Au}]
\end{align}
These would be the nonlinear waveguide coefficients as seen by a quasi-cw pulse. 
Let us mention that in the standard non-vectorial notation where the longitudinal field vanishes, the denominator of Eq. (\ref{eq:Gamma-app}) contains a factor $n_0^2$. We here include a similar factor $\tilde n^2$, which as we show below is a generalized effective index that can be calculated in the specific simple case we investigate here, the strip waveguide. This factor is in principle absent in the vectorial versions of the literature, e.g. in \cite[Eq. (17)]{Marini2013}, simply because it is buried in the connection between the electric field and the magnetic field, and when the longitudinal component is nonzero the relation between the electric field and the power cannot  generally be parametrized in this simple way. On the other hand, in the special case investigated here, the infinite strip waveguide has only $e_y$, $e_z$ and $h_x$ nonzero modes.\footnote{As opposed to the infinite strip waveguide, the finite strip waveguide has a small nonzero $e_x$ contribution, which we neglect; it does not appear in this simple analysis we present here and requires solving the problem using a finite-element mode-solver.} This means that $\int_{dA} ({\rm Re}[\vec{e}\times \vec{h}^*]\cdot \hat z)^2=\frac{k_0^2}{\beta^2}\int_{dA}({\rm Re}[\varepsilon])^2 |e_y|^4=\frac{k_0^2}{\beta^2}\bar n^4 \int_{dA}|e_y|^4=\tilde n^2 \int_{dA}|e_y|^4$, where we have defined
\begin{align}
\bar n^4&=\frac{\int _{dA}({\rm  Re}[\varepsilon])^2|e_y|^4}{\int _{dA } |e_y|^4}\\
\label{eq:ntilde}
\tilde n^2&=\frac{\bar n^4}{n_{\rm eff}^2}= 
\frac{\int _{dA}({\rm  Re}[\varepsilon])^2|e_y|^4}{n_{\rm eff}^2\int _{dA } |e_y|^4}
\end{align}
i.e. $\tilde n$ is a kind of generalized effective index. 
This simple case also means that the vectorial version of the nonlinear field fractions simplify
\begin{subequations}
\label{eq:theta-simple}
\begin{align}
\theta_{\rm Au}&=\frac{4}{3}\frac{\int_{dA_{\rm Au}} |\vec{e}|^4} 
{\int_{dA} |e_y|^4}
\\
\theta_{\rm TaO_5}&=\frac{1}{3}\frac{\int_{dA_{\rm TaO_5}} (2|\vec{e}|^4+|\vec{e}^2|^2)} 
{\int_{dA}|e_y|^4}
\\
\theta_{\rm SiO_2}&=
\frac{1}{3}\frac{\int_{dA_{\rm SiO_2}} (2|\vec{e}|^4+|\vec{e}^2|^2)} 
{\int_{dA} |e_y|^4}
\end{align}
\end{subequations}

\begin{figure*}[bt]
\centering
\includegraphics[height=3.4cm]{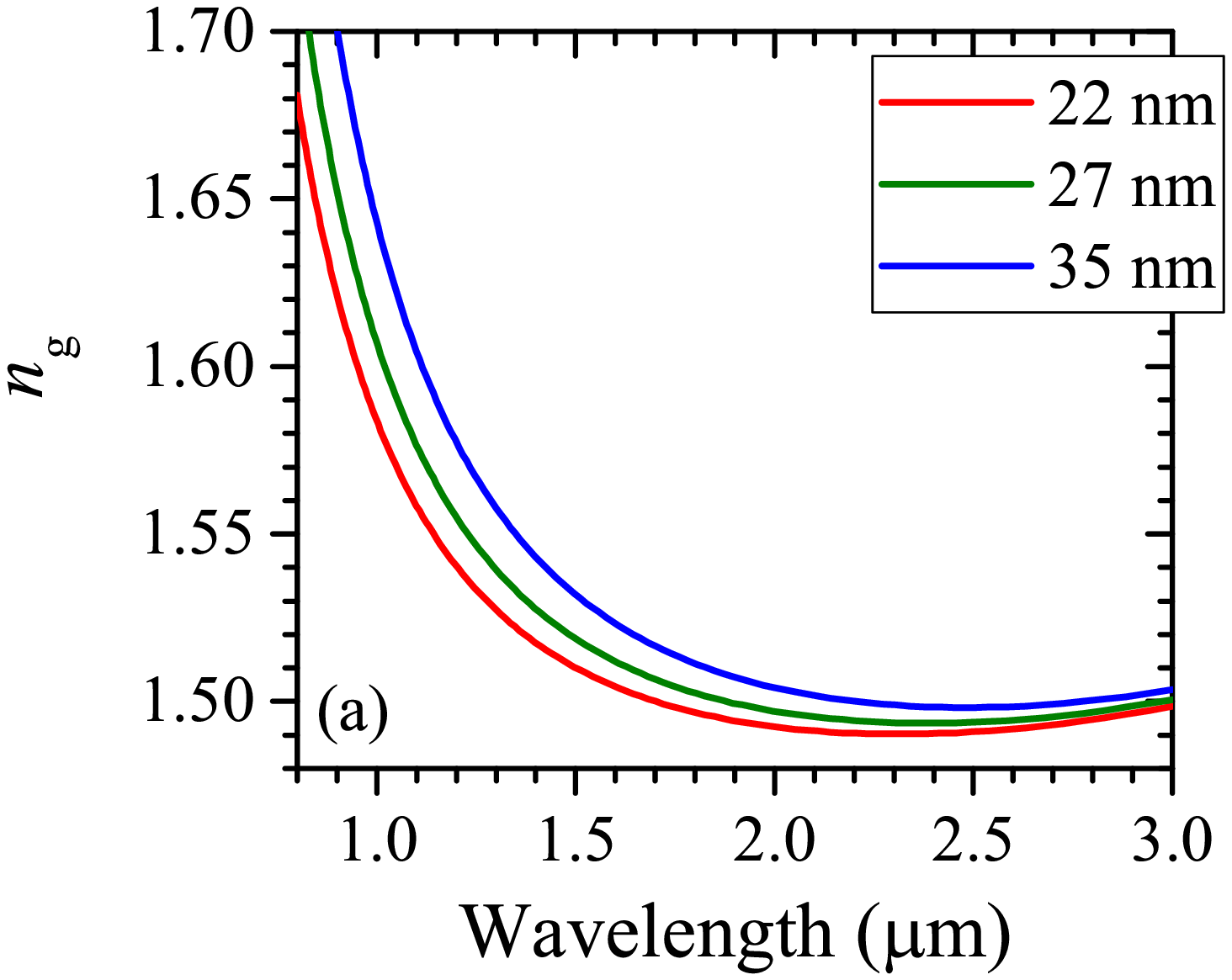}
\includegraphics[height=3.3cm]{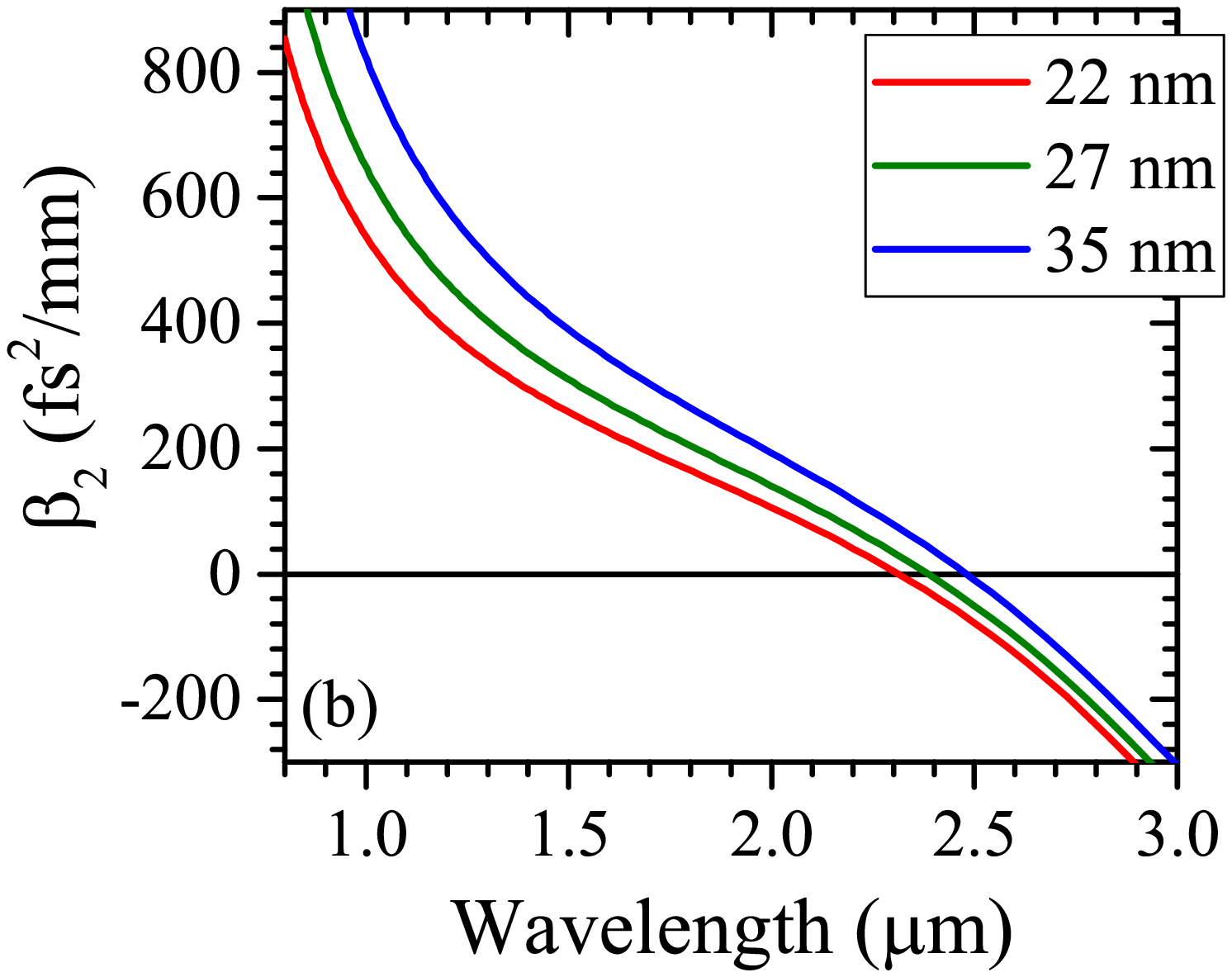}
\includegraphics[height=3.4cm]{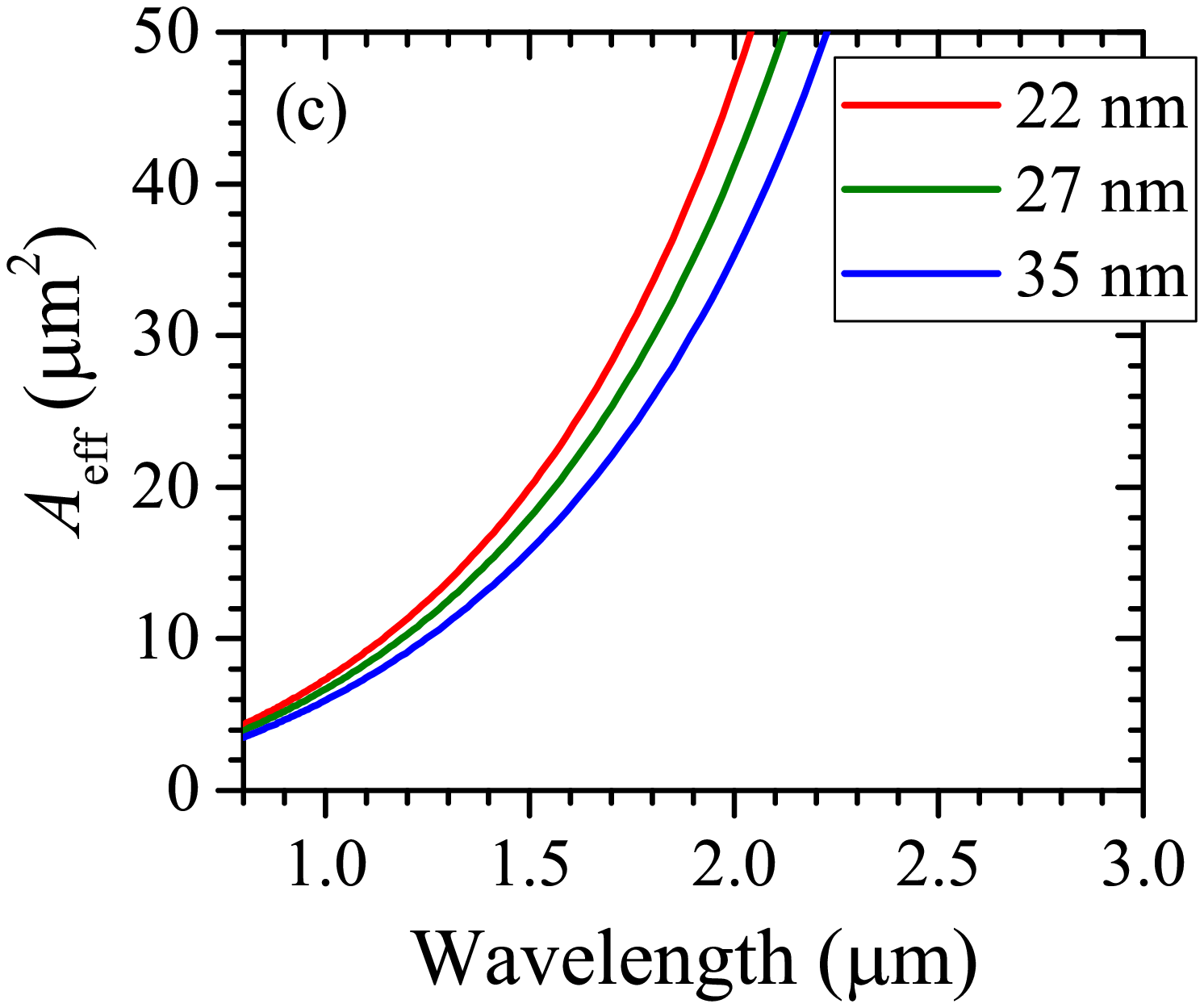}
\includegraphics[height=3.3cm]{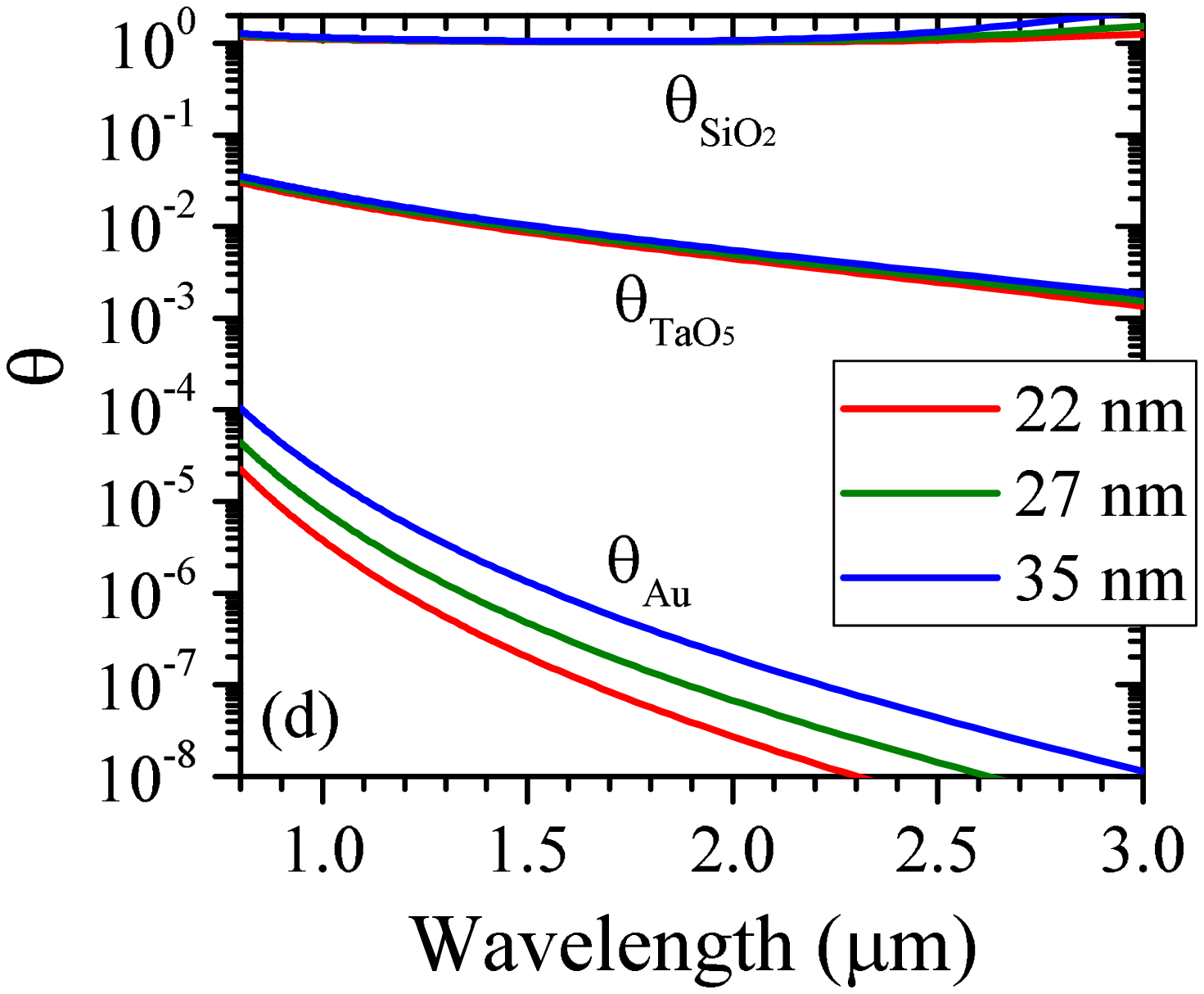}

\caption{Variation of the fundamental mode waveguide parameters vs wavelength in the vectorial calculations. (a) Group index $n_g=c\beta_1$, (b) GVD $\beta_2$, (c) effective mode area $A_{\rm eff}$ and (d) nonlinear field fractions $\theta_j$. 
}
\label{fig:dispersion}
\end{figure*}

We found that the vectorial calculations gave nonlinear coefficients up to an order of magnitude higher than the scalar calculations, where only $e_y$ was used. A summary of the results of the calculations is presented in Fig. \ref{fig:dispersion}. 

From Eq. (\ref{eq:NLSE_Au}) we can express the envelope in a general way separating conveniently the power and the phase as $A(\zeta,\tau)=\sqrt{P(\zeta,\tau)}e^{i\phi(\zeta,\tau)}$. The imaginary part of the resulting equation gives the nonlinear power loss equation
\begin{align}\label{eq:NLSE-int-Au}
\frac{\partial P}{\partial\zeta}+\alpha P
+\beta_{\rm NL}^{\rm cw} P\int_{-\infty}^\infty d\tau' h_T(\tau-\tau') P(\zeta,\tau')=0
\end{align}
where we neglected dispersion (this is a good approximation in the mm-scale waveguides we use) and self-steepening. In order to be able to reduce this to an ODE that we can solve, we note that for a pulse much longer than the characteristic response time of $h_T$ this reduces to 
\begin{align}\label{eq:NLSE-int-long-pulse-limit}
\frac{\partial P}{\partial \zeta}+\alpha P+\beta_{\rm NL}^{\rm cw} P^2=0
\end{align}
We see that the $\beta_{\rm NL}^{\rm cw}$ value here has the role of being equivalent to the nonlinear absorption parameter found in the long-pulse cw limit. In order to reduce Eq. (\ref{eq:NLSE-int-Au}) to a form like (\ref{eq:NLSE-int-long-pulse-limit}) we need to estimate the convolution to get rid of the time dependence. What we suggested in \cite{Lysenko2016b} was that to a good approximation
\begin{align}\label{eq:NLSE-int-Au-betap}
\beta_{\rm NL}^{\rm cw} P(\zeta,\tau)\int_{-\infty}^\infty d\tau' h_T(\tau-\tau') P(\zeta,\tau')\\
\nonumber \simeq
\beta_{\rm NL}(T_0)P^2(\zeta,\tau)
\nonumber \equiv
\beta_{\rm NL}^{\rm cw}\rho(T_0)P^2(\zeta,\tau)
\end{align}
where $\beta_{\rm NL}(T_0)$ is a corrected value of $\beta_{\rm NL}^{\rm cw}$ that depends on the pulse duration $T_0$ of the pump through the convolution. To account for this we introduced the so-called "correction factor" $\rho$. This is a dimensionless quantity that gauges the strength of the overlap  
\begin{align}
\label{eq:rho_semianalytical}
\rho(T_0)={\rm max}_\tau [p(\tau)\int_{-\infty}^\infty d\tau' h_T(\tau-\tau')p(\tau')]
\end{align}
where $p(\tau)=P(\zeta=0,\tau)/P_0$ is the normalized temporal dependence of the input power with a characteristic pulse duration $T_0$, e.g. $p(\tau)=\exp(-\tau^2/T_0^2)$ for a Gaussian pulse. 

The main message is that even in the short-pulse limit we can with a good approximation write
\begin{align}\label{eq:NLSE-int-long-pulse-limit-rho}
\frac{\partial P}{\partial \zeta}+\alpha P+\rho(T_0)\beta_{\rm NL}^{\rm cw} P^2=0,
\end{align}
which can be solved analytically. Thus, the only difference is that the correction factor is applied to the nonlinear term, which has the limits $\rho\rightarrow 0$ for $T_0\rightarrow 0$ and $\rho\rightarrow 1$ for $T_0\rightarrow \infty$. Of course, the fact that the TTM predicts an absence of nonlinear effects for an extremely short pulse is a consequence of the fact that the nonlinearity is purely delayed with no instantaneous contributions.  

\bibliographystyle{unsrt}
\bibliography{literature}

\end{document}